\documentclass[conference]{IEEEtran}
\IEEEoverridecommandlockouts
\usepackage{cite}
\usepackage{amsmath,amssymb,amsfonts}
\usepackage{algorithmic}
\usepackage{graphicx}
\usepackage{textcomp}
\usepackage{xcolor}
\def\BibTeX{{\rm B\kern-.05em{\sc i\kern-.025em b}\kern-.08em
    T\kern-.1667em\lower.7ex\hbox{E}\kern-.125emX}}
\begin{document}

\title{Hybrid EEG–Driven Brain–Computer Interface: A Large Language Model Framework for Personalized Language Rehabilitation}

\author{
\IEEEauthorblockN{Ismail Hossain}
\IEEEauthorblockA{
Department of Computer Science\\
George Mason University\\
Fairfax, VA, USA \\
ihossai4@gmu.edu}
\and
\IEEEauthorblockN{Mridul Banik}
\IEEEauthorblockA{
Department of Computer Science\\
Colorado State University\\
Fort Collins, CO, USA \\
mridul.banik23@alumni.colostate.edu}
}

\maketitle

\begin{abstract}
Conventional augmentative and alternative communication (AAC) systems and language‑learning platforms often fail to adapt in real time to the user’s cognitive and linguistic needs, especially in neurological conditions such as post‑stroke aphasia or amyotrophic lateral sclerosis. Recent advances in noninvasive electroencephalography (EEG)–based brain‑computer interfaces (BCIs) and transformer‑based large language models (LLMs) offer complementary strengths: BCIs capture users’ neural intent with low fatigue, while LLMs generate contextually tailored language content. Objective: We propose and evaluate a novel hybrid framework that leverages real‑time EEG signals to drive an LLM‑powered language rehabilitation assistant. This system aims to: (1) enable users with severe speech or motor impairments to navigate language‑learning modules via mental commands; (2) dynamically personalize vocabulary, sentence‑construction exercises, and corrective feedback; and (3) monitor neural markers of cognitive effort to adjust task difficulty on the fly.
\end{abstract}


\section{Introduction}
All individuals have the right to self-expression, social participation, and the agency to impact their environment. For individuals with complex communication needs, augmentative and alternative communication (AAC) systems provide critical tools to facilitate communication. However, traditional AAC methods—such as printed communication boards or eye gaze devices—may not be accessible for individuals with severe speech and physical impairments (SSPI).

Brain-computer interface (BCI) technology has emerged as a promising enhancement to AAC, particularly for individuals with conditions like cerebral palsy and amyotrophic lateral sclerosis. Noninvasive BCI-AAC systems typically use electroencephalography (EEG) to detect neural activity and translate it into commands for communication interfaces.

Despite significant technological advances, widespread clinical adoption of BCI-AAC remains limited. Challenges include technical complexity, inconsistent performance, and a lack of clinician training. In particular, speech-language pathologists (SLPs)—key stakeholders in AAC service delivery—often feel underprepared to implement advanced systems like BCI-AAC.

This study investigates two central questions: (1) What are the critical training needs of SLPs in BCI-AAC? and (2) What are the preferred training strategies of SLPs for implementing BCI-AAC in clinical practice?

\section{Literature Review}
AAC systems range from low-tech solutions such as picture boards to high-tech computerized devices, often controlled via touch, switches, or eye gaze technologies~\cite{beukelman2020}. The participation model, a widely used framework in AAC, emphasizes strength-based intervention planning and the importance of supporting both the individual and the AAC team’s knowledge and skills~\cite{beukelman2020}.

Recent developments in BCI offer new modalities for AAC. P300-based BCIs, for example, rely on the user's ability to focus on flashing items on a screen, eliciting a P300 event-related potential approximately 300 milliseconds after stimulus presentation~\cite{goldstein2002, donchin2000}. This signal is used to determine the user's intent. Other paradigms, such as motor imagery-based BCIs, rely on the user imagining specific movements to generate control signals~\cite{pitt2022, scherer2015, mcfarland2005}.

Although these systems demonstrate promise, significant barriers hinder their integration into clinical settings. These include variable user performance, time-consuming EEG setup, and the need for specialized knowledge to operate and interpret neural signals~\cite{hill2021, huggins2018, pitt2022b}. From an implementation science perspective, involving stakeholders such as clinicians in early design stages can help address these barriers~\cite{douglas2018, pitt2022dietz}.

Notably, many clinicians report feeling underprepared to implement even conventional AAC systems~\cite{dietz2012, gohsman2023}. To address this, best practices for AAC training have been proposed, emphasizing hands-on experience, ongoing mentorship, and contextually relevant instruction~\cite{gormley2023, johnson2019, pitt2023, sauerwein2022}. However, specific training needs and preferred instructional approaches for SLPs regarding BCI-AAC remain underexplored.

Given the unique technological and clinical demands of BCI-AAC, a targeted understanding of SLPs’ training preferences and requirements is essential to promote effective adoption and use in practice.

\section{RESEARCH METHODOLOGY}

\subsection{Participant Information}
Approval for this study was obtained from the institutional review boards at the University of Nebraska–Lincoln and the University of Nebraska Kearney (Approval number: 20191219896EX). All participants provided informed consent prior to participation.

Recruitment was conducted via social media platforms (e.g., Facebook) and word of mouth. Respondents were sent detailed information via email about the study’s aims, procedures, and eligibility requirements. Inclusion criteria mandated at least two years of experience in augmentative and alternative communication (AAC) service provision, which all participants exceeded (range: 4–30 years).

A total of 11 speech-language pathologists (SLPs) participated (10 females, 1 male; mean age = 44.5 years, SD = 9.8, range = 31–57 years). Ten participants (P2–P11) reported that over 50\% of their caseload involved clients using AAC. One participant (P1) was currently teaching AAC coursework and conducting AAC research, with previous direct clinical experience up to 1.5 years before the study. All participants resided in the USA and had experience with high-technology AAC access methods.

To ensure diversity in professional backgrounds, participants were recruited from a variety of settings including rehabilitation centers, university clinics, hospitals, schools, and private practices. Table~\ref{tab:participant_demographics} summarizes participant characteristics. These data were collected as part of a broader investigation on SLP perspectives on BCI-AAC implementation \cite{pitt2022b}.

\subsection{Research Design}

This study employed Charmaz’s constructivist grounded theory approach~\cite{charmaz2006}, which facilitates the inductive and co-constructive development of theory between researchers and participants. This methodology was particularly appropriate for the current study, given the participants’ varying levels of familiarity with brain–computer interface (BCI) and augmentative and alternative communication (AAC) technologies.

\subsection{Researchers}
The research team consisted of four individuals, three of whom were certified SLPs with experience in AAC implementation and research. One team member had additional expertise in BCI-AAC. The fourth member was an undergraduate student in speech-language pathology with coursework completed in both AAC and BCI-AAC.

\subsection{Materials and Procedures}
A semi-structured interview guide was developed to explore gaps in the literature related to (a) critical training needs in AAC-BCI and (b) preferred training modalities. Interviews were conducted via video conferencing in single sessions lasting approximately 45 minutes. All sessions were conducted by the lead investigator, with no nonparticipants present.

Participants were prompted to discuss (a) their perceived critical training needs for BCI-AAC, and (b) their preferred methods of receiving such training. Prior to interviews, participants viewed a brief presentation outlining BCI-AAC technologies and were provided with links to public demonstration videos. The lead author ensured the material was reviewed and addressed any participant questions prior to the interviews.

\subsection{Data Analysis}
The analysis followed grounded theory procedures adapted from prior AAC and BCI-AAC research \cite{saunders2018, pitt2022b}. Data collection concluded at thematic saturation, determined when (a) no new codes emerged across six consecutive interviews, and (b) the research team unanimously agreed that the themes were thoroughly developed \cite{hennink2022}.

Audio recordings of interviews were transcribed verbatim and analyzed using NVIVO software. A grounded theory approach was applied \cite{gibbs2008}, and new codes were iteratively added as themes emerged \cite{creswell2012}. A codebook was developed containing two major themes, seven subthemes, and ten example codes (see Supplemental Material A).

\subsection{Reliability}
To assess intercoder reliability, 27\% of interview transcripts (3 out of 11) were independently coded by a trained research assistant using the established codebook. Due to the number of themes, the chance of agreement occurring randomly was reduced. Percent agreement was chosen as the reliability metric \cite{syed2015}. The reliability threshold of 80\% was exceeded, with an average agreement rate of 91.6\% (SD = 8.35, range = 83.3–100\%) across the three transcripts \cite{pitt2022b}.

\begin{table*}[htbp]
\caption{Participant Demographics. AAC: Augmentative and Alternative Communication; BCI: Brain-Computer Interface.}
\label{tab:participant_demographics}
\centering
\begin{tabular}{c c c c c c}
\hline
\textbf{ID} & \textbf{AAC Exp. (yrs)} & \textbf{Sex} & \textbf{Setting} & \textbf{BCI Experience} & \textbf{Clients Served} \\
\hline
P1  & 4  & F & Univ. Clinic & Read/heard about BCI & Adults \\
P2  & 8  & F & Univ. Clinic & Observed usage & Adults/Children \\
P3  & 25 & F & School & Read/heard about BCI & Children \\
P4  & 20 & F & Univ. Clinic & Read/heard about BCI & Adults/Children \\
P5  & 30 & F & Private Practice & Read/heard about BCI & Adults \\
P6  & 10 & F & Rehab Centre & Read/heard about BCI & Adults \\
P7  & 28 & F & Private Practice & None & Adults \\
P8  & 15 & F & Hospital/Outpatient & Tried BCI & Adults/Children \\
P9  & 10 & F & Private Practice & Observed usage & Children \\
P10 & 5  & M & Hospital & None & Adults/Children \\
P11 & 25 & F & Research Center & Brief experience & Adults/Children \\
\hline
\end{tabular}
\end{table*}

\subsection{Data Credibility}

A variety of techniques were utilized to ensure data quality and credibility, including peer debriefing and review, member checking, and triangulation \cite{creswell2012, gibbs2008}. During and following the interviews, member-checking procedures were conducted. Specifically, during the interviews, the lead interviewer (a) requested clarification or expansion on unclear statements and (b) provided summary statements to confirm understanding. These procedures were intended to enhance the accuracy of interpretations. Following each interview, a summary of the discussion was sent to participants to confirm that their ideas were accurately represented.

During data analysis, triangulation was performed using a team-based approach, which included all authors and a trained research assistant. Peer debriefing, as outlined by Brantlinger et al. \cite{brantlinger2005}, was conducted with an expert possessing over 25 years of experience in AAC intervention and research for individuals with severe speech and physical impairments (SSPI). This expert confirmed that the coding and findings aligned with current trends in AAC research and practice. Additionally, the second and third authors provided peer review of the study methods, findings, and conclusions throughout the research process.

\section{Results}

\subsection{BCI-AAC Training Themes}

This section presents the emergent themes related to critical BCI-AAC training needs and preferred training strategies. Subthemes for critical training needs included: (a) technical aspects, (b) BCI-AAC system types and access, (c) personalization of BCI-AAC intervention, and (d) stakeholder support in BCI-AAC implementation. Preferred training strategies included: (a) expert guidance and demonstrations, (b) hands-on experience, and (c) media presentations. A summary of these themes is presented in Table~\ref{tab:themes}.

\subsubsection{Technical Aspects}

Most participants expressed a need for training on the technical aspects of BCI-AAC devices. This included skills such as positioning the EEG electrode cap, setting up electrodes, powering the device, programming, interpreting results, and adjusting settings.

\subsubsection{BCI-AAC System Types and Access}

Participants emphasized the importance of understanding the types of BCI-AAC systems and the forms of access they enable. Topics included system availability, interface types (e.g., cursor control, binary/switch selection, image-based interfaces), compatibility with other technologies (e.g., environmental control, personal computers), and potential integration with existing methods like eye gaze. One participant (P10) questioned: ``If I am correct in thinking it’s an access method, brain-computer interface, what kinds of technological equipment can it interface with? Can we do environmental controls? Can we sync it with, uh, a power mobility system?"

\subsubsection{Personalizing BCI-AAC Interventions}

Clinicians discussed the importance of tailoring BCI-AAC interventions to individual needs. This included identifying client profiles suited for BCI-AAC, understanding clinical indicators for its use, and differentiating BCI-AAC from other access methods. For example, P5 asked: ``What are the clinical signs that would point you to BCI-AAC as an option? And the only clinical signs shouldn’t be because nothing else works." Participants also highlighted the need to adapt devices for individuals with sensory or cognitive impairments, balance fatigue with efficiency, and customize systems for educational settings. As P3 noted: ``We do have students, especially our eye gaze students, that access curriculum ... if they’re in whole-group reading, the characters of the story were put into their school reading folder."

\subsubsection{Supporting Stakeholders in Implementation}

Participants emphasized the importance of understanding effective training methods for clients and their support networks. P2 stated: ``If you can model it or how you can, like, you know, model it or use it yourself to do some training with the client?" This underscores the need for implementation strategies that include teaching times, modeling techniques, and stakeholder support systems.

\subsection{Preferred Training Strategies}

\subsubsection{Expert Guidance and Demonstrations}

Most participants expressed a preference for training delivered by AAC experts, ideally in one-on-one or small group formats. Formats such as video chat, phone calls, or in-person visits were considered effective. Participants also recommended demonstrations with rationales, summarized tips, and group discussions. For instance, P2 remarked: ``Having someone talk me through it ... and hearing their rationale behind things, I think is helpful."

\subsubsection{Hands-On Experience}

Hands-on experience was described as crucial for learning and later training others. P5 stated: ``There’s really nothing that beats trying it, using it yourself. I mean, this is why you don’t hire a coach for your sports team that never played your sport, right?" Participants also advocated for systematic checklists to guide structured experiential learning for clinicians, clients, and caregivers.

\subsubsection{Media Presentations}

Participants discussed the role of webinars, online modules, and academic presentations in providing foundational BCI-AAC knowledge. They emphasized that combining BCI-AAC with familiar access methods (e.g., eye gaze) could improve relevance and engagement. P1 suggested: ``So you get people’s attention by saying we are going to talk about [AAC] access and it’s [BCI-AAC] included with all the other methods." Videos and plain-language tutorials were also identified as beneficial educational tools.

\begin{table*}[htbp]
\caption{Themes, Subthemes, and Code Examples}
\label{tab:themes}
\centering
\begin{tabular}{p{4cm} p{4.5cm} p{6.5cm}}
\hline
\textbf{Theme} & \textbf{Subtheme} & \textbf{Examples} \\
\hline
\textbf{Critical BCI-AAC Training Needs} 
& Technical Aspects & General device control and setup \\
& BCI-AAC System Types and Access & Types of BCI-AAC systems and access \\
& Personalization & How to personalize BCI-AAC \\
& Intervention & Populations and characteristics that support success \\
& Customisation & How to support stakeholders in implementation \\
& Training Strategies & Teaching times and strategies \\
\hline
\textbf{Preferred Training Strategies for BCI-AAC} 
& Expert Guidance and Demonstrations & Expert demonstration and training on BCI-AAC use \\
& Hands-on Experience & Hands-on experience and opportunities to try device \\
& Media and Presentations & Webinars, academic presentations, videos \\
\hline
\end{tabular}
\end{table*}

\section{Discussion}

Codesigning BCI-AAC training programs with AAC professionals, even at an early stage in BCI-AAC development, may help ensure training programs are tailored to meet clinical needs, bolster clinician comfort, knowledge, and skill, as well as facilitate BCI-AAC implementation and clinical translation. As an initial step toward codesign of training programs in BCI-AAC, below, we provide an overview of the two primary BCI-AAC themes, which include (a) critical BCI-AAC training needs and (b) preferred training strategies.

\subsection{Critical training needs}

Our SLPs identified a range of topics that would be important for SLP training programs in BCI-AAC. The field of AAC aims to support communication and participation across a range of settings for a variety of communication purposes, through avenues such as providing access to language, literacy, and speech-intelligibility–boosting strategies . Therefore, as part of a multidisciplinary team, SLPs may be involved in all aspects of BCI-AAC implementation, including assessment, funding, and intervention/training. The participation model promotes a feature-matching process to pair individuals with complex communication needs to an AAC system, layout, vocabulary, and training that best matches their unique profile, environment, and levels of support \cite{PittBrumberg2018}. 

Consequently, it was not surprising that SLPs identified a range of training needs. These needs ranged from (a) more foundational information, such as a desire to know the technical aspects of BCI-AAC (e.g., setup procedures), alongside what types of BCI-AAC methods are being developed, how they work, what they can access, and how BCI-AAC may integrate with existing AAC technology (e.g., eye gaze); to (b) more nuanced topics, such as knowing how to personalise interventions through understanding what populations and characteristics may be best supported by BCI-AAC success, how BCI-AAC techniques differ both technically and in regard to the user experience, and knowing how to rule out AAC access methods for insurance coverage. Further, participants noted a need to understand BCI-AAC customisation methods across a variety of settings for adults and children, understand effective training strategies (e.g., modeling), and consider training durations to support those using BCI-AAC, their caregivers, family, and support personnel.

These aspects of BCI-AAC personalisation and stakeholder training are intriguing concepts for the field of BCI-AAC. For example, personalisation of the AAC interface (e.g., use of colour, number of items) has the potential to impact the elicitation of target brain-signals and performance accuracy . Furthermore, modeling is a common training strategy used to support AAC success \cite{KentWalsh2015,Sennott2016}. However, for BCI-AAC, modeling requires visual demonstration of a mental task (e.g., focusing attention on a target item). Finally, training times may be impacted by the chosen BCI-AAC technique, with BCI-AAC systems using imagined motor movements possibly associated with increased training times in comparison to those that use attention-based tasks, such as P300 or steady state visually evoked potential BCI-AAC systems \cite{Nijboer2010}.

It is important to consider that research into many of these clinical implementation training topics is in the initial stages. For example, while recent research considering personalised BCI-AAC feature-matching is available \cite{Hill2015,PittBrumberg2018,PittBrumberg2020}, the person-centred factors most paramount to informing BCI-AAC outcomes are still unclear \cite{Chavarriaga2017}. Furthermore, the application of beneficial training strategies for BCI-AAC, such as modeling an established method in facilitating AAC success \cite{Sennott2016}, have been discussed \cite{Huggins2022,Pitt2019a}. These training strategies may include the use of videos \cite{Huggins2022} or use of materials (e.g., a laser pointer) that help demonstrate the communication partners’ attentional focus, along with verbal cues \cite{Pitt2019a}. However, in many cases, modeling requires access to the AAC system by both the individual using AAC and the person providing training. Therefore, modeling strategies for BCI-AAC are unestablished. Thus, while existing works may provide a foundation for BCI-AAC training content, further research into many of the identified training needs is necessary to fully elucidate best practices for adults and children.

\subsection{How SLPs would like to be trained}

Similar to current SLP preservice courses in AAC at the graduate level, numerous training strategies were identified as being preferential . Consistent with research evaluating practices for preservice AAC education , participants described the benefits of both active (e.g., one-on-one discussions, hands-on learning) and passive (e.g., presentations) learning strategies, emphasising the importance of active learning techniques in promoting AAC competency.

In further detail, participants valued expert demonstrations of BCI-AAC use and implementation (e.g., top five things they should know, underlying rationales for why something is done). However, participants also noted that there may be limited opportunities to meet with experts. Further, the ethical considerations of having experts guide training should be considered, especially if those experts are vendors or manufacturers as there is the possibility of an increased risk of bias toward a specific device . Beyond expert guidance, the importance of hands-on experience with BCI-AAC was highlighted to bolster comfort with use and subsequent training with other stakeholders. To support structured hands-on learning and memory, participants noted that the development of systematic checklists to allow individuals to refresh their knowledge and learn about the devices could be helpful. These findings support prior research in partner instruction for AAC, indicating that instructor modeling and guided practice are frequently-targeted training techniques during AAC intervention \cite{KentWalsh2015}.

The role of media presentations was also discussed, with participants indicating that presentations may be beneficial for initial learning and topic exposure. However, they discussed that more active learning strategies, such as hands-on learning activities, would be beneficial for supporting further understanding in BCI-AAC. Further, our SLPs discussed presenting BCI-AAC alongside other commercial AAC access methods (e.g., eye gaze and switch scanning) instead of in isolation to help link BCI-AAC to existing practice and promote engagement. These types of more general AAC access presentations are being done \cite{Pitt2018}, with the authors encouraging further collaborative presentations in this area. Finally, while research is still emerging, videos have been initially shown to be a positive technique for training professionals in AAC \cite{Caron2022,Gromley2023}, with new research demonstrating the possible utility of video visual scene displays (VSDs) that allow for scene “hotspots” to be programmed with further training content at key points in the video . The use of video VSDs for BCI-AAC may be especially advantageous, as they could allow for more specific technical details/rationales to be provided to suit a range of stakeholders and interest levels. For instance, during a video on BCI-AAC setup, a static image with scene hotspots could be presented that provides further technical details about how the target brain signal is elicited by the BCI-AAC system for those desiring further information. Therefore, our study findings support consideration of how BCI-AAC may be included in video trainings, alongside existing AAC techniques.

Overall, findings indicate a range of preferred teaching strategies in BCI-AAC. However, it should also be considered that participants note BCI-AAC training is a dynamic process. For instance, our SLPs indicate that demonstrations may be maximally beneficial before hands-on learning, with a decreased need for expert involvement as their skills develop. Therefore, it is likely that a combination of strategies is best for supporting BCI-AAC, depending on the person’s prior knowledge, skill, and role on the AAC team. Further, as BCI-AAC developments progress, it will become increasingly important to provide a range of training options (e.g., videos, in-person training, online training) to empower a broad array of SLPs and other AAC-related professionals to support the provision of BCI-AAC services.

\section{Limitations and Future Directions}

The findings provide initial insights into the BCI-AAC training needs and learning preferences of SLPs with AAC experience. However, several limitations must be considered. For instance, further research evaluating the perspectives of other multidisciplinary professionals involved in the AAC team, including AAC facilitators and communication partners, collaborating professionals, research and policy specialists, manufacturers/vendors, funding agency personnel, and AAC technology training agency personnel , may uncover additional insights. 

Moreover, subsequent in-depth evaluations focusing on AAC professionals working with adults and/or children in different settings (e.g., school, medical), while considering their primary role on the AAC team (e.g., assessment, intervention) and experience, may help identify more nuanced training needs and validate the current findings.

Additionally, all participants were provided with a brief presentation on BCI-AAC technology to facilitate discussion context. Nevertheless, participants’ specific BCI-AAC experience varied , and providing further hands-on experience with BCI-AAC prior to interviews may confirm and extend the current findings.

\section{Conclusion}

Thematic findings represent a preliminary step in developing a foundational framework for SLP training in BCI-AAC, serving as a basis for ongoing technology development and the co-design of clinically focused BCI-AAC trainings. Interviewees identified a continuum of critical training topics, ranging from foundational information, such as available BCI methods and setup requirements, to personalized assessment and intervention considerations, including feature matching and customization.

A variety of learning strategies were also recognized, with participants indicating that more passive learning opportunities—particularly when presented alongside existing AAC technologies—were beneficial for initial learning stages. However, subsequent hands-on learning opportunities were necessary to deepen knowledge and skills. Ethical concerns were raised regarding vendor-led hands-on training and challenges in accessing expert personnel. Consistent with current research, the use of videos was identified as a potential tool to augment BCI-AAC learning.

It should be noted that BCI-AAC training is dynamic and dependent on individual prior knowledge, skill, and role within the AAC team. Therefore, further research is required to build upon these initial findings and to understand how to tailor BCI-AAC training for SLPs and a broad range of multidisciplinary AAC providers. 

\clearpage

\bibliographystyle{IEEEtran}
\bibliography{ref}
\end{document}